# Conceptual Modeling of Events Based on One-Category Ontology

Sabah Al-Fedaghi
*salfedaghi@yahoo.com, sabah.alfedaghi@ku.edu.kw*
Computer Engineering Department, Kuwait University, Kuwait

**Summary**
In previous works, we proposed a one-category (entitled thimac) conceptual model called a *thinging machine* (TM), which integrates staticity (e.g., objects) and dynamism (e.g., events) without losing valuable aspects of diagrammatic intuition in conceptual modeling. We proposed applying TM to conceptual modeling in software engineering (e.g., on or above the level of UML as a conceptual modeling language). In this paper, to show such an application in software engineering, we first present a complete high-level description of a library service system to demonstrate the TM's applicability. Furthermore, we explore the TM's features, emphasizing the realization of thimacs as *events*. The purpose is to develop better understanding of the TM notions by contrasting them with their uses in related fields. The notion of an event plays a prominent role in many fields of study, including philosophy, linguistics, literary theory, probability theory, artificial intelligence, physics, and history. A TM event is *a static thimac with a time breath (time subthimac)* that infuses dynamism into the thimac. It arises from how the TM static region is infected with time. Such a view is contrasted with some philosophical and linguistics definitions of an event (e.g., unit of experience – Whitehead). We also raise interesting issues (e.g., event movement) in this study.

## 1. Introduction

As software engineering must often interact with the outside world, it needs tools to develop a description of domain fragments in which software must be developed. Such a venture necessarily encounters general philosophical problems of describing the world.

Conceptual modeling entails developing an abstract model with appropriate simplification of reality [1]. The model involves an explanation of the real system, which is capable of producing all possible input–output behaviors and integrating various components of a system to be refined into a more concrete executable model. In short, the conceptual model defines what and how it is to be represented [1]. Two types of conceptual models are typically identified: a domain-oriented model that provides a detailed representation of the domain and a design-oriented model that describes the model's requirements in detail [1]. This paper focuses mainly on domain-oriented conceptual modeling.

Modeling is deemed more of an "art" than a "science" [2]; therefore, according to Karagöz [3],

> [I]t is generally assumed defining methodical ways to develop conceptual models is difficult. The evolution of newer engineering fields, such as systems and software engineering, has shown that using well-defined modeling notations, following defined processes, and utilizing software tools definitely improve effectiveness. [3]

### 1.1 Sample Current Approach

For example, during the last twenty years, many studies have promoted the utilization of object-oriented language UML for conceptual modeling [3] [4]. According to Breiner et al. [4], UML has value only in a software project's early stages and is discarded in the project's later stages through implementation and testing. Additionally, according to Breiner et al. [4], the UML approach suffers from the difficulty of learning and applying 14 types of diagrams and ensuring consistency across the diagrams [4]. Breiner et al. [4] also claimed that

> UML diagrams arose from a variety of needs and applications, and were not designed to work together. Its wide variety of constructions overlap, so that it is often unclear what type of model should be used to capture a particular observation. The underlying semantics for UML modeling was an afterthought, defined after the fact and rarely called on in practice.

As a possible solution, Breiner et al. [4] proposed diagrammatic models that "look very similar to UML class diagrams" and are grounded in the mathematical category theory.

### 1.2 Alternative Conceptual Constructs

*Conceptual* modeling attempts to model a system based on *concepts*. Developing such models touches the psychological and abstraction realms. This effort requires supplying unambiguous categorization with elements of discreteness and hierarchically ordered representations of the modeled domain. Conceptual construct meanings have to be defined carefully using ontology to analyze and enrich the capacity to capture knowledge about an application domain. Often, however, rigorous definitions of these constructs are missing.

Categorization is the elementary task for the construction of our understanding of the world, through which ideas are recognized, differentiated, and understood. Categorization represents the "the most basic phenomenon of cognition" [5]. Historically, categorization has been intended to enumerate everything that can be expressed without composition or structure [5]. The representation of the modeled domain is formed utilizing natural language and diagrams with semantics determined by mental and nonmental factors. Conceptual representation may also require inviting and/or uncovering new categorizations.

Conceptual categories are fundamental to the mapping between a model and its domain. Bradley and Bailey [6] commented that it is hard to say much about a category under which *every thing* falls; nonetheless, candidates are available for such a category, including *thing*, *entity*, and especially, *object*. According to Sinha and Gärdenfors [7], we are naively accustomed to thinking of objects as the most fundamental ontological category of the physical world.

One difficulty related to the notion of "object" is that the world's variety seems to lie not only in the assortment of its objects "but also in the sort of things that happen to or are performed by them" [8]. There would seem to be a mode difference between objects and *events* that are said to occur, happen, or take place [8]. Additionally, objects have relatively crisp spatial boundaries and vague temporal boundaries, and events have vague spatial boundaries and crisp temporal boundaries.

According to Casati et al. [8], some philosophers simply would deny the conceptual distinction between events and objects and would treat the distinction as one of degree: a thing is "a monotonous event; an event is an unstable thing." Some philosophers claim that, although objects and events are featured as "the basic units from which to build a descriptive system," the primacy of objects is strongly supported by phenomenological considerations (see sources in Casati et al. [8]).

### 1.3 Objective: Advocating Thimacs as a Base for Conceptual Modeling

In this paper, we claim that the subtle difference between objects' and events' fundamental conceptual constructs plays an important role in constructing conceptual models. In previous works (e.g., [9]), we proposed a one-category conceptual model (called a *thinging machine* [TM]) that integrates staticity (e.g., objects) and dynamism (e.g., events) without losing valuable aspects of diagrammatic intuition in conceptual modeling. We applied a TM for conceptual modeling in software engineering (e.g., on or above the UML level as a conceptual modeling language).

In this paper, we further explore the thimac notions, emphasizing the realization of *thimacs as events*. The purpose is to advocate this approach by developing a better understanding of the TM notions by contrasting them with their uses in related fields. As we will argue, a TM event is *a static thimac that has a time breath (time subthimac)* that infuses dynamism in the thimac. The event arises from how the TM static region is infected with time. We contrast such a view with philosophical and linguistic definitions of an event (unit of experience – Whitehead [10]).

### 2. TM Modeling

The TM model is a conceptualization of how things/machines can be merged into a complex of interrelated thimacs (i.e., things that are simultaneously machines). The thing and the corresponding machine "exist" as one thimac; the thing reflects the unity and the machine shows the structural components, including potential (static) actions of behavior. Behavior refers to sequence of events or, in other woods, the occurrence of actions.

A thimac is a thing. The thing is what can be created (appear, observed), processed (changed), released, transferred, and/or received. As will be discussed later, a thing is manifested (can be recognized as a unity) and related to the whole TM or as a static (timeless) phenomenon. Later, when we discuss dynamism, this thing becomes an "instance" when supplemented with time (which is also a thing) to form a dynamic unity of a thing called an *event*. Thus, things are at the TM static description (model) and are at the dynamic model when merged with time.

The thimac is also a machine that creates, processes, releases, transfers, and/or receives. Fig. 1 shows a general picture of a TM. The figure indicates a "field" with five "seeds" of potentialities of dynamism. Aristotle identifies matter with potentiality, for example, wood, as matter, has the potential to be a statue. Potentiality is an Aristotelian notion. For example, what we call "create" at the static description is a potentiality of creation. According to Aristotle, a thing has within it "a starting-point of change in another thing or in itself insofar as it is other" (see *Stanford Encyclopedia of Philosophy - Aristotle's Metaphysics*). This is what we call static action. The create TM static action is the thing's capacity to be a gate for the create variation (difference) when interacting with time.

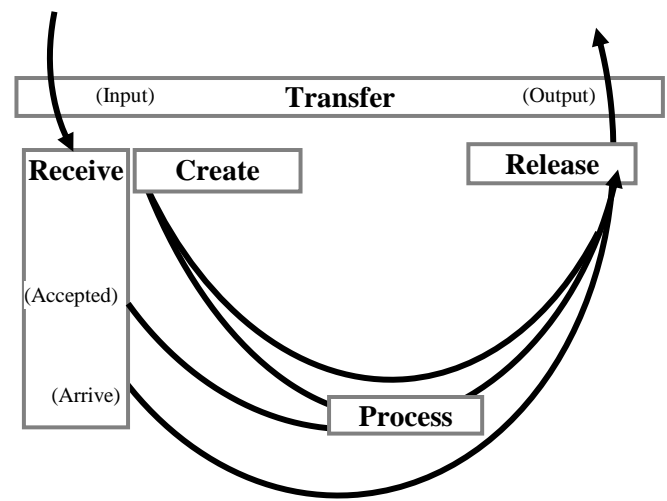

**Fig. 1 A thinging machine.**

Aristotle thought that potentiality is indefinable. Aristotle noted, "What is awake is in relation to what is asleep, and what is seeing is in relation to what has its eyes closed but has sight" (see *Stanford Encyclopedia of Philosophy—Aristotle's Metaphysics*). For example, according to Aristotle, the wood has (at least) two different potentialities because it is potentially a table and potentially a bowl. TM contains five types of these potentialities, creation (e.g., a bowl), processing (e.g., decorated bowl), release, transfer, and receive (e.g., a bowl moves from one person to another). In TM, there are no other potentialities (e.g., sell, change, display, give, clean, break, etc.); all of these actions can be expressed in terms of the five generic actions.

The five generic of actions can be described as follows.
- The appearance (coming into existence in the system) of a new thing (create)
- The variation (change) in the same thing (called process)
- The movement from one field to another (release, transfer, and receive)

*Appearance* is the phenomenon of becoming or "existing" within the system; *variation* is a change in the same thing; and *movement* occurs among machines. They are "seeds" of potential actions.

A TM structure may be viewed as a thing when considered as a whole; therefore, the thing may flow to yet another machine, as Fig. 2 shows. Here, a structure refers to the discrete composite components of the TM description, including the multilevel net of TM machines.

### 2.1 Preliminary Notes about the Notion of Action Potentiality

Initially, an *action* is a state of readiness to become an *event*. Prior to an event's emergence, an action in its potentiality state is part of a static description. An action as part of a static state (which seems at first contradictory) is just a static object in the common meaning. What moves (changes) an object that has the capacity of motion (change) is the *event* that emerges from the (potential) *action* and *time* "possess" (as in spirit possession) each other. Therefore, a verb in a sentence does not refer to an (actual) activity unless the sentence represents an event. Usually, natural language ambiguity blurs such a difference between potential action and actual action (event).

Returning to conceptualizing the notions of action and event, we find that, for Aristotle, an action is a type of event "with an inherent end" [11]. For Aristotle, an action is a "potentiality" (a capacity for action), and time ignites "actuality" (a type of event that means the existence of the thing [11]). Note that Aristotle did not explicitly include events in his categories [11]. Aristotle's event-related analysis is based on linguistic forms in which verbs are viewed as dynamic beings. This linguistic approach continued in various works.

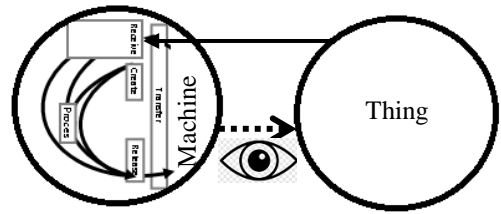

**Fig. 2 A machine may be viewed as a thing that flows to a machine.**

In recent times, this linguistic-based conceptualization can be seen in Davison's analysis of sentences searching for event structures [11].

TM is based on recognizing generic actions. TM "beings" (i.e., thimacs) have, in themselves, a principle of motion (generic actions). From Aristotle, we take the notion of "potentiality"; that is, actions need a time element to exhibit dynamism. Rest is actualized by time change with no internal change. In a TM, actuality is the fulfillment of generic actions that potentially exist.

### 2.2 The Thimac

A thimac forms an arena in which a potentiality acts on a thing that happens to be there according to its position in one of the five seeds of potential action. It is a static field in the sense that there is no dynamism. Similar to the concept of "field" in the physical sciences, the TM field is a region in which potentiality acts on (is applied to) things but in five ways. When this static field is joined by a time field (thimac), each thing is stimulated according to its seed (i.e., create, process, release, transfer, and receive). We call this combination of static and time fields a *generic event*.

A TM may be viewed as a thing when considered as a whole; therefore, the thing may flow to yet another machine. Thimacs are created, and they create their subthimacs. Thimacs first have to be created so they can create, process, and move things. Any thimac "exists" (in the TM diagram), so it, in turn, creates other thimacs. For simplicity's sake, we consider the presence of a box as a sign that the thimac exists.

Thimacs are conceptual (mind-made) fields of potential action seeds developed to make sense of the world. A thimac exhibits sufficient "togetherness" to form a bounded whole of subthimacs. Therefore, a thimac is a generally mechanistic ontology in which we see a thing that is conceptualized as the mereological totalities of subthimacs.

All things are created, processed, and transported (acted on), and all machines (thimacs) create, process, and transport other things. Things "live" or "pass through" other machines. The thing is a presentation of any "existing" (appearing) entity that can be "counted as one" and is coherent as a unity. A noun is usually used as a label for things and what we perceive and can identify, even if we have no words to name it.

Machines house other things and provide pathways for their flow. The unity of thing and machine forms a thimac. In such a blend, a single thimac is a fusion of two manifestations, flow and machines, for other flowing things. The actions in the machine are ordered in a specific way (Fig. 1). As Fig. 1 shows, a TM can be viewed as a coordinated system of flow. The flow is not a link type (e.g., a class of relationship in ER); rather, it is a transformation from one potentiality of action to another.

### 2.3 The TM

Fig. 1 can be described in terms of the following generic (has no more primitive action) actions.
**Arrive**: A thing moves to a machine.
**Accept**: A thing enters the machine. For simplification, we assume that all arriving things are accepted; hence, we can combine the arrive and accept stages into one stage: the **receive** stage.
**Release**: A thing is ready for transfer outside the machine.
**Process**: A thing is changed, but no new thing results.
**Create**: A new thing is born (found/manifested) in the machine and is realized from the moment a thing arises (emergence) in a thimac. Things come into being in the model by "being found." Creation in metaphysics involves bringing entities from a state of nonbeing into existence. The TM model limits this creation to appearance in the model. *Create x* in a model means "there is" *x*. After the instance of creation, the entity may move toward processed or released, or it may stay in the creation state.
**Transfer**: A thing is input into or output from a machine.

Additionally, the TM model includes the triggering mechanism (denoted by a dashed arrow in this study's figures), which initiates a flow from one machine to another. Multiple machines can interact with each other through the movement of things or through triggering. Triggering is a transformation from one series of movements to another.

### 3. Complete Example of TM Modeling

In this example, (from [12]) the librarian can list the library books. From there, a book may be selected for addition, or a new book may be created. Both of these use cases include a list of the related book copies (see Fig. 3). The librarian is also able to list the books, and he/she may select related authors. Fig. 4 shows the TM model for this subset of a librarian's use cases in a library system.

We will model this example in a TM with some modifications related to the general understanding of the case. For example, related books can be extracted from the main list of all books and not updated separately, as Cruz [12] seems to indicate.

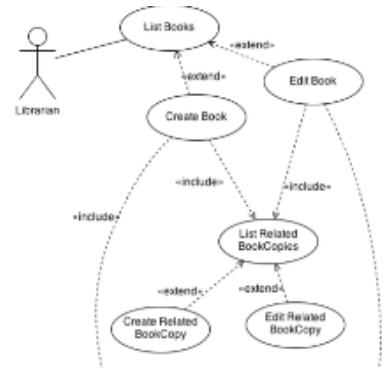

Fig. 3. The library system's use case model (partial, from [12]).

We can also produce the related books and authors by inputting the book ID without tying such a list to new or updated books. Additionally, for simplification, we ignore the case of book copies. Some boxes are eliminated for simplification. There are two first-level machines: the librarian and the library system. The box with the broad oval at the top of the library system can be labeled a "books list" machine; however, we opt not to label it because its function is clear: the list of books is sent and updated (created) either to the librarian or to the library system.

### 3.1 Description of the TM Static Model

The following are shown in Fig. 4.
- The librarian requests access to the library system (circle 1 in the figure), and such a request flows to the system (2), where it is processed (3). Note that all generic actions (create, release, transfer, receive, and process) are potentialities that reflect dynamism. The request is a thing, and its machine spread across the librarian and the system.
- We can consider the sequence of potentialities that extend between circles 1–3 as the conceptual field construct of "the librarian accesses the library system."
- This is our understanding of this type of interaction between the librarian and the system. Of course, such an interaction needs data to be realized, but regardless of the data's type and size, the flow between circles 1 and 3 remains the same.
- Assuming that the process determines that the request is acceptable, it triggers (4) the download of the book list (5) to the librarian (6). Additionally, with the list, the system prompts (dashed red line to the left of the figure) the librarian to determine whether he/she intends to add a new book or edit data about an existing book (the downward vertical red dashed arrow in the librarian machine). Accordingly, the librarian makes (7) his/her selection, which goes (8) to the system, where it is processed (9).
- If the librarian selects "new," the system creates a book record to be filled (10) and sends it to the librarian (11), who supplies data to create a new filled book record (12) and sends it to the system (13). The system sends (14) that record to be added to the book list, which requires the retrieval of the current list (15) and the new record (16) to be processed (17) for creating a new list (18).

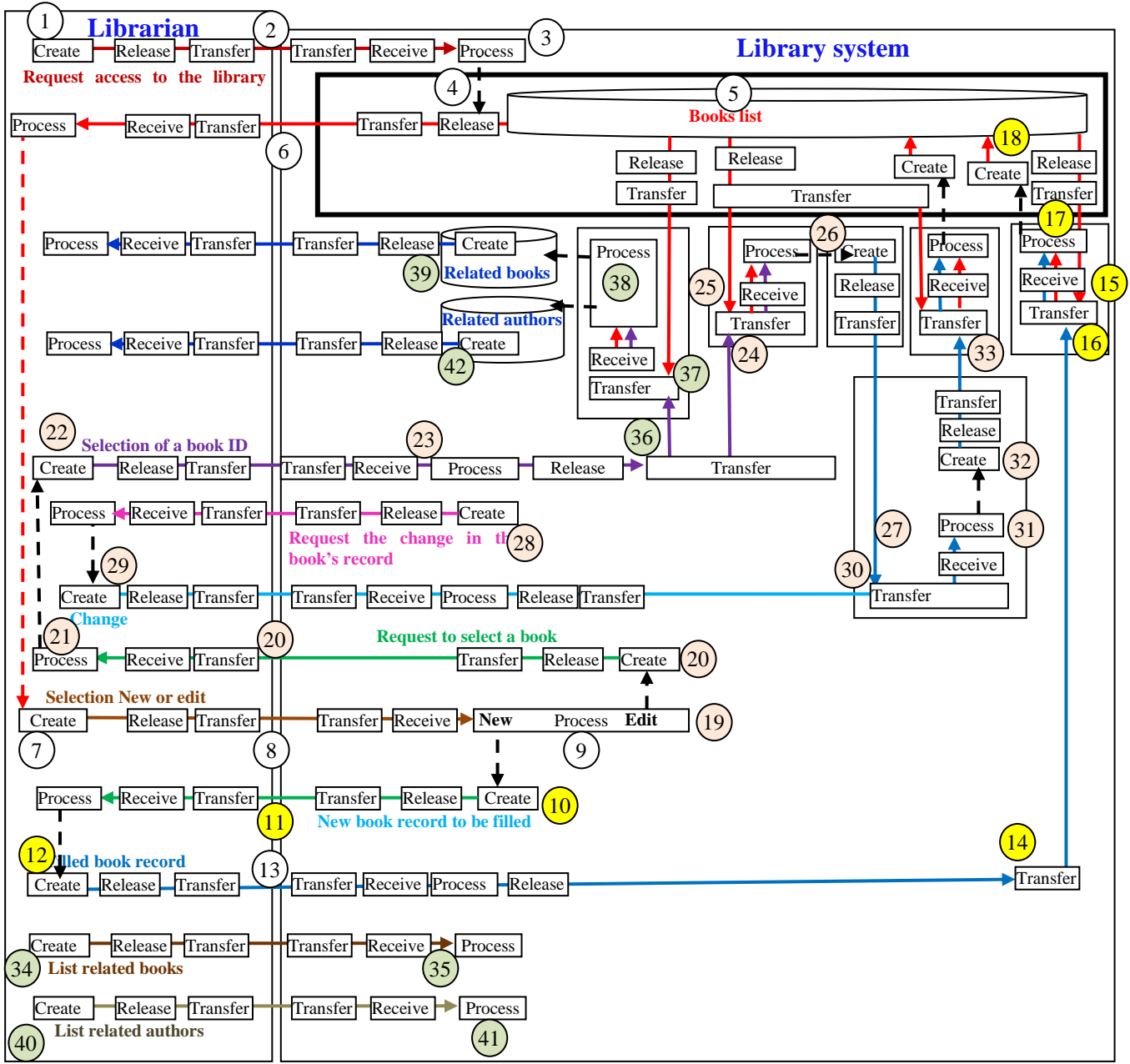

**Fig. 4 The library system's TM static model.**

- If the librarian chooses to edit a book record (19), then the system creates a request (20) to select a certain book from the previously downloaded list (6). The request flows to the librarian (21), who processes the request (20) and sends the book ID (22). The system receives the book ID (23) and sends it as a procedure (24) that compares it with IDs of records in the book list (25), thus finding the input key record (26). Note that this is indicated by a process that triggers the creation (appearance in the global view of the system) of the required record (26). The book's found record flows to edit (27).

- Editing the book's record requires (28) the librarian to input the changes (29—e.g., change publication date). Hence, in the relevant procedure, the book record (28) and the change (30) are processed (31) to create a new record (32), which (33) updates the current list of books as before (18 and 19).

- Now, we come to the part that is different from Cruz's [12] description due to a lack of a complete understanding on our part and to simplify the example. However, the TM model can be extended to accommodate any other parts that we do not cover. We assume that the librarian can list books and the authors for any given book separately from the operations of adding a new book or editing a current book. Therefore, (at 34 and 35, bottom left corner), the librarian requests a list of all books related to a certain book.

- He/she can identify the book as before (23 and 24). Hence, the book ID (23 and 24) is processed with the current list of books (36, 37, and 38) to produce related books (39). A similar procedure is followed to produce the list of relevant authors (40, 41, and 42).

## 3.2 The Behavioral Model: Events

The TM's behavioral model is constructed as the chronology of events in the modeled system. An event is defined as a combination of a subdiagram of the TM static (standing still) model (i.e., the event region) plus a time subthimac, which *activates* (come alive, and thus trigger change) the region. An event must involve at least one generic action over some stretch of time. Time in this description is the thimac of being dynamic (motion: flow of things), analogous to the life of a physical body. Time flows (transfer, receive, etc.) in the "totality" machine that includes the static and time thimacs (Thus, there is no need for the notion of "super time" in time flows). Additionally, time is defined with a non-empty static TM subdiagram (thus, avoiding the problem of time applied to a time subdiagram). Motion is described as actuality of the unity mentioned above (static thimac + time). Another way to say that is that time induces change (generic actions); however, because there is no "super-time," no changing of change appear in the model.

Fig. 5 shows the event *The librarian requested a list of authors*.

For simplification purposes, the event may be represented by its region. As shown in Fig. 6, we identify the following events. The figure is simplified by denoting action sequences by their first letters. For example, CRTTRP denotes the sequence of actions: create, release, transfer, transfer, receive, and process.

$E_1$: The librarian requests access to the system.
$E_2$: The system downloads the book list, giving the librarian the choice of either starting a new book or editing a current book.
$E_3$: The librarian makes a selection from new/edit options.
$E_4$: The system processes the librarian's selection and recognizes the selection of a new book.
$E_5$: The system processes the librarian's selection and recognizes the selection of an edited book.
$E_6$: The system sends a request to fill the record of a new book.
$E_7$: The librarian supplies the new book data, and then the system receives the new book data and sends it to update the book list.
$E_8$: The current book list is downloaded to update it.
$E_9$: The book list and the new book record are processed to add the new book record.
$E_{10}$: The new book list replaces the old one as the latest list.
$E_{11}$: The system requests the edited book ID from the librarian.
$E_{12}$: The librarian supplies the book ID that is received by the system and used to retrieve the book record.
$E_{13}$: The book record is retrieved from the book list and then sent to the librarian for editing.
$E_{14}$: The books list and the book ID are processed to retrieve the book record from the list.
$E_{15}$: The book record is extracted from the list and sent to the librarian for editing. Note that C (create) in the event denotes the appearance of the book record as an independent entity.
$E_{16}$: The system requests the changes to be made for the book record.
$E_{17}$: The librarian sends the changes that are received by the system to be used to update the book record.
$E_{18}$: The book record and the changes are processed to update the record.
$E_{19}$: A new record that includes the changes is created and sent to the librarian to update the current list. Note that C (create) in the event denotes the appearance of the new version of the record as a new entity.
$E_{20}$: The book list and the new record are processed to replace the old version of the record.
$E_{21}$: A new version of the list is created that replaces the old version.
$E_{22}$: The librarian requests a book list related to a given book.
$E_{23}$: The librarian requests an author list related to a given author.
$E_{24}$: The book list is retrieved and sent to select relevant books and authors.
$E_{25}$: The book list is processed according to the given ID, and relevant books and authors are selected.
$E_{26}$: The relevant books are sent to the librarian.
$E_{27}$: The relevant authors are sent to the librarian.

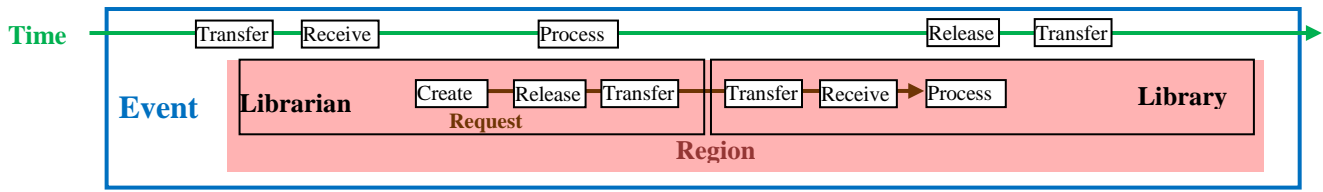

**Fig. 5 The event *The librarian requested a list of authors*.**

Fig. 7 shows the library behavioral model in terms of the chronology of events.

The resultant static, dynamic, and behavioral models can be used to build information and control systems for the library services.

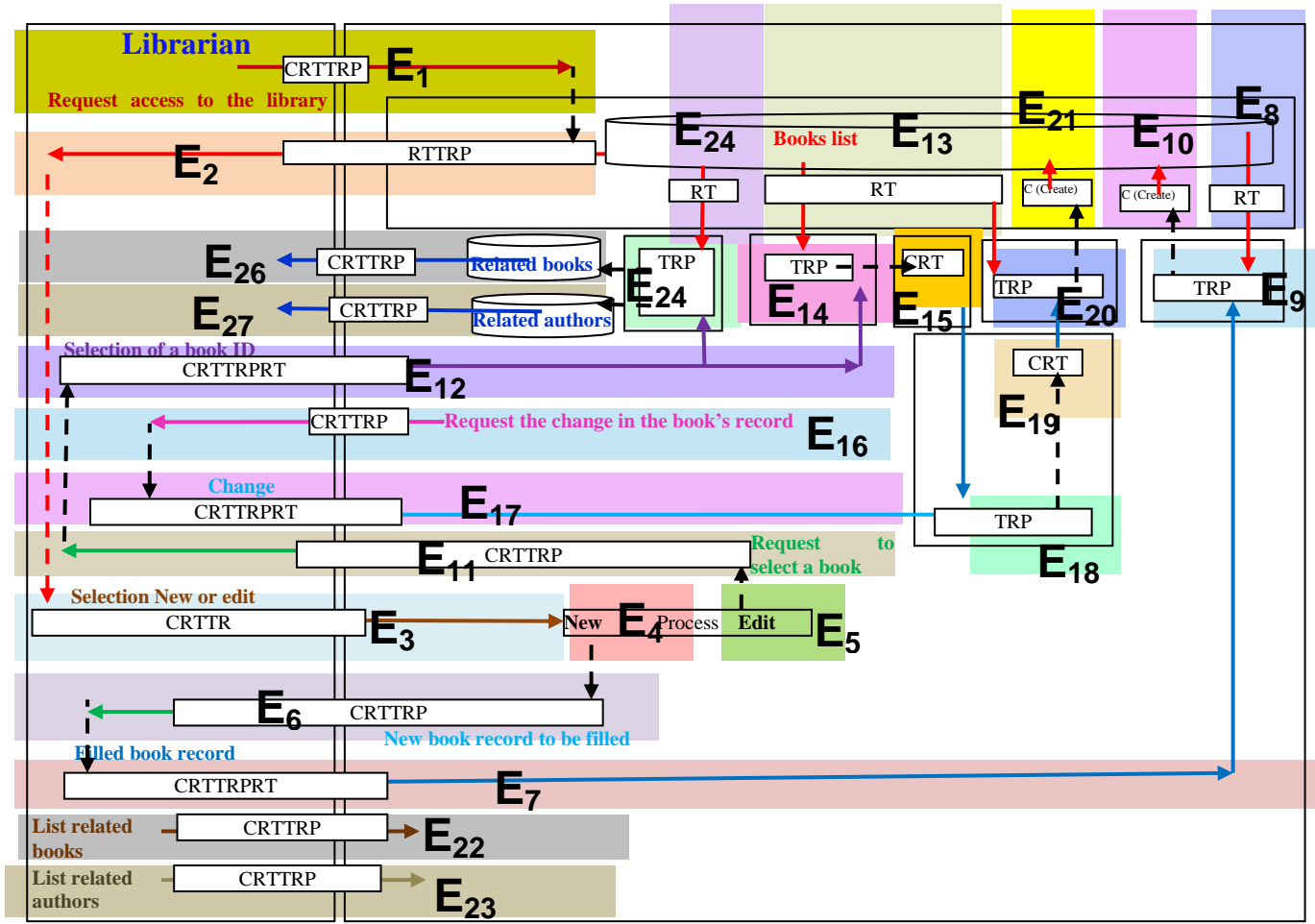

**Fig. 6 The events model of the library system.**

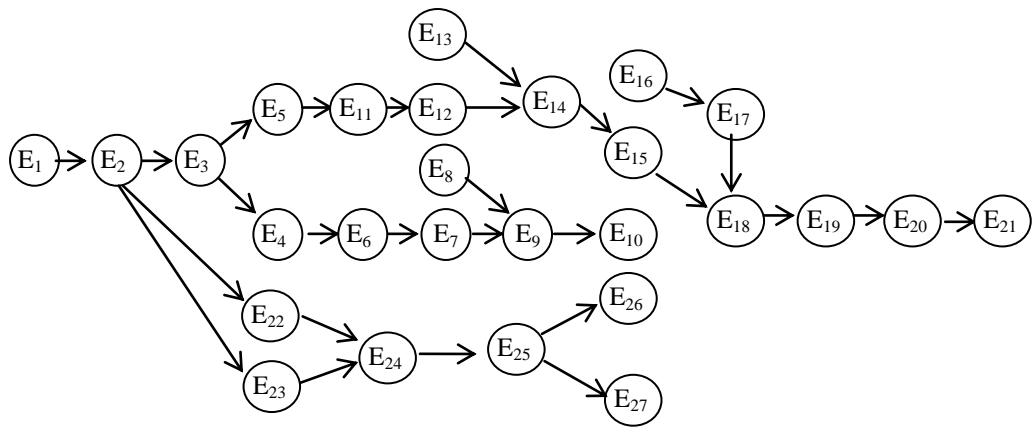

**Fig. 7 The behavior model of the library system.**

## 4. Illustration Potential Actions and Events in the TM

To illustrate the relationship among actions and events, assume that our domain includes just two things, X and Y, which are shown in Fig. 8. X is the machine, and Y is the thing that is created by Y. Fig. 8 (left) is a picture of a *static* situation of potential action. Fig. 8 (right) shows another *static* picture, in which Y is now in the *process* stage of X. The TM static diagram is the union of all possible static situations or, in other words, the sum of the regions specified as a TM diagram. The totality of what the static diagram represents is a thimac, a topological construction from machines and submachines. During the flow of a thing, the thing may be in any stage in the TM diagram. In each situation, the stage (potential action) is part of the *static* situation. A *change* occurs when time is involved.

It is clear that the five actions—create, process, release, transfer, and receive—are not the so-called states. In Fig. 8, a thing in a stage does not change. For example, if the thing changes from a green to a blue color, this change occurs *inside* the *process* stage. When a thing moves from, say, the create stage to the release stage, the relevant change is in the context of the thing, not in the thing itself (e.g., position, orientation, etc.). To see the involved changes in the presence of time, consider a typical two-state system as a typical binary signal model. Let us describe it in terms of a two-level height, as shown in Fig. 9. Each stage and time form an event, as shown in Fig. 9.

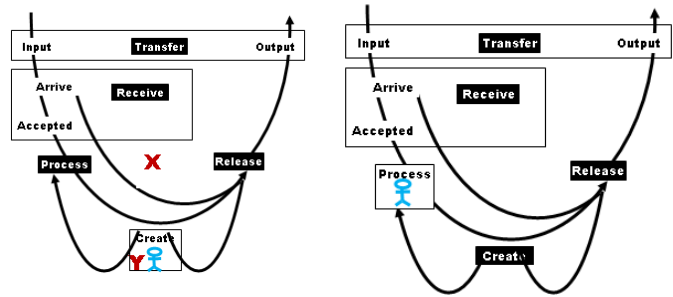

**Fig. 8 Machine X and thing Y**

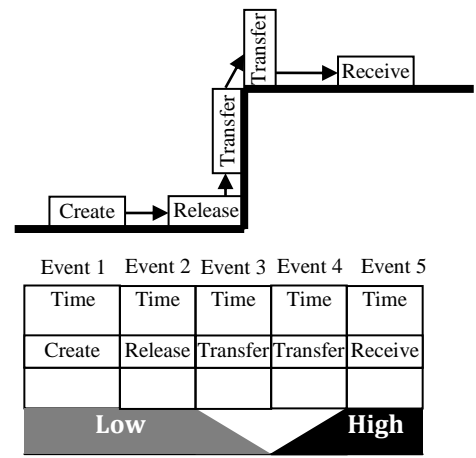

**Fig. 9 Two-state model and TM states**

As can be observed, the two states of low and high are not generic events, whereas create, release, transfer, and receive are generic. Linking the various stages of the thing's path requires viewing a thing (e.g., signal) at different periods as a single thing. Fig. 9 shows what we previously called a potential action "possessed" by "time durations" to generate events. The "fuzzy" events of transfer are an interesting phenomenon that needs more analysis to understand the nature of events.

## 5. A Glimpse on the Event Notion

The focus in the remaining part of this paper will be on exploring the notion of an event as manifested in different applications and relating it to TM events.

### 5.1 General

The notion of an event plays a prominent role in many fields of study, including philosophy, linguistics, literary theory, probability theory, artificial intelligence, physics, and—of course—history [8]. Event perception, event recognition, event memory, event conceptualization, and segmentation have long been studied in several fields of psychological research [8].

In cognitive science, the perceived world is structured into objects, places, and actions that form parts of events [16], as well as numbers [17] as core knowledge domains, which form the framework of perceptual categories.

Common sense typically construes events as "concrete, dated particulars, i.e., as non-repeatable entities with a specific location and duration" [8]. The structure of language attests to the primacy of the event in human cognition. Event structure (i.e., the combination of constituents encoding objects, actions, and location) is the fundamental building block for sentence meaning and grammar [7].

Whitehead [10] recognized that "the event is the ultimate unit of natural occurrence." Whitehead defined events as chunks in the life of nature that refer to the experience of activity (or passage) [13]. According to Shipley [14], "Events appear to be a fundamental unit of experience, perhaps even the atoms of consciousness, and thus should be the natural unit of analysis for most psychological domains."

In common language, the term "event" encompasses wider range meanings, including things that happen on short or long timescales, such as interactions between subatomic particles or the orbit of Saturn around the Sun [15]. The linguist's use of the notion of an event may not adhere with the vision scientists whom themselves have "changed their use and understanding of such notions over the years" [8].

Gärdenfors [16] suggested that events are an overarching category for combining different perceptual categories and combining objects, actions, and locations. Event structures are represented in terms of conceptual spaces—one for actions and one for results—and mappings between these spaces.

### 5.2 Verbs and Events

From the linguistic point of view, the TM's five generic actions imply the reduction of all verbs to five generic verbs. According to Tversky et al. [18], verbs do not describe components of *events* the way nouns alone can. Consider, for example, the list of verbs: *take*, *spread*, *fold*, and *put*. Without knowledge of the *objects* being acted upon, we cannot know if this is about baking a cake or putting away the laundry. This implies that verbs are not parts of the world (next to objects); rather, they are components (alongside with generic actions) in determining the structure of events.

TM introduces a different picture (see Figs. 10 and 11) and divides the linguistic expression into two levels. In the TM static level, objects and verbs (specified as generic actions) form a "structure" that becomes a dynamic description of events when time is added to the structure. Fig. 10 shows the TM representation of these verb sequences: *take*, *spread*, *fold*, and *put*. The figure indicates that machine B takes a thing from machine A, performs two types of processing (spreading and folding), and puts the thing in machine C. According to Tversky et al. [18], folding flour into a batter and folding a sheet are achieved with very different body movements. By contrast, in the TM, assuming that this refers to hands that are part of a human arm, machine B's "hands" perform the same movement, diagrammatically, for different objects (e.g., a sheet and clay).

The problem stems from verb genericity. For example, *take* is a nongeneric verb because it can be expressed in the generic verbs *release* and *transfer*. It is not possible to take anything without the holder of the thing releasing and letting it go (output) and the receiver getting (input) and receiving it. *Take*, *spread*, *fold*, and *put* are not generic actions. For example, in the case of *take*, assume there are two agents (A and B in Fig. 11); hence, the first agent releases, and the second agent receives, as well as the first agent output (transfer) and the second agent input (transfer).

We also emphasize here that, in TM, verbs (generic actions) and objects (things) form a static structure, and that events are the dynamism of this structure when time is involved. A generic action (e.g., create) is a potentiality in the static structure, similar to the way sense as a flow (an action) is represented by an arrow in the structure. The static description, as a stable all-encompassing frame of potentialities, does not specify individuals such as instances or events. In the static form (e.g., TM diagram/subdiagram), everything is there; nothing corresponds to time (past, present, or future); and nothing corresponds to, say, the principle of noncontradiction. However, what is "there" is loaded with potentiality that can be exemplified by actuality.

Additionally, these types of linguistic studies mix staticity with dynamism. Fig. 11 expresses the behavioral model that corresponds to Fig. 10. Here, the verbs *take*, *spread*, *fold*, and *put* take their form as events that integrate their regions of the TM static description and time.

### 5.3 Events and Generic Events

According to a recent article [19], Davidson [20] showed (1967) that the same event may be compositionally described by multiple modifiers (e.g., *Jones buttered the toast,* and *Jones buttered the toast slowly*). Such an analysis type views a sentence as a whole "lump-sum" and mixes static representation with its corresponding dynamic semantics. TM representation converts the sentence into its generic actions and then identifies dynamic features in terms of events. For example, Fig. 12 shows the TM representation of *Jones buttered the toast*, and Fig. 13 shows the corresponding event representation. Fig. 14 shows the logical sequence of events.

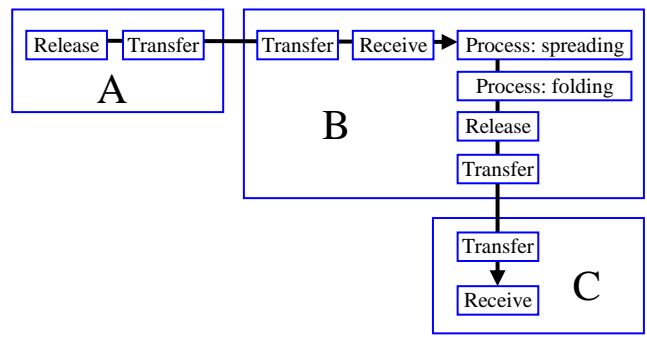

**Fig. 10 TM static representation of** *take*, *spread*, *fold,* **and** *put*

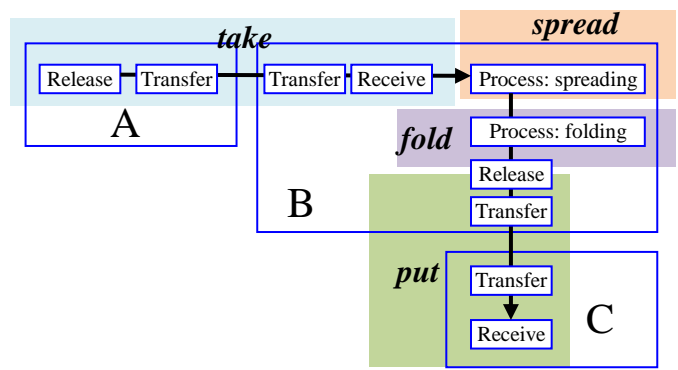

**Fig. 11 TM representation of the events** *take*, *spread*, *fold,* **and** *put*

The events in Fig. 14 reflect a high-level abstraction of the elementary events shown in Fig. 13. Each of these elementary events takes time (change). Assuming that Jones handles the toast before handling the butter, we can develop movie clips of the events, as shown in Fig. 15. First, Jones appears (created); the toast appears, gapped by Jones (transfer, transfer, and receive); the butter appears (created), gapped by Jones (transfer, transfer, and receive); and the toast is buttered. A few of these scenes are illustrated in Fig. 16.

Events overlap, creating events at varying levels of granularity (may be called compound events [14]; e.g., a metal sphere is simultaneously rotating and getting warm), and then its rotation and its getting warm appear to be simultaneous distinct (generic) events within the same thimac (see discussion in Casati et al. [8]).

This phenomenon of hierarchical and overlapping events can be viewed as a mechanism of Gestalt grouping: The ongoing activity stream is parsed into meaningful wholes [15]. All events, generic or at a higher level, are made of the same stuff: the five generic actions and time. High-level events form a coarse description, whereas generic events are the finest level of event segmentation. The events may also have other subthimacs of associated properties (e.g., intensity).

## 6 Events and Movement

A TM event is *a static thimac with a time breath (time subthimac)* that infuses dynamism in the thimac. It arises from how the TM static region is infected with time. Such a view is not far from the linguistics definition and structure of an event that consists of three parts [21]: a predicate (e.g., TM subdiagram), an interval of time on which the predicate occurs (TM time subthimac), and a situation under which the predicate occurs (TM). However, the TM event cannot be described only as a unit of experience (apprehending being [10]); rather, it is made up of multilevel units of dynamic phenomenon based on, at the lower level, the five actions as units that are grasped by our experience. Dynamism is a regulating mechanism of the static form that aligns with reality through such machinery as igniting and chronologizing actions, logicalizing, and executing/controlling processes. An event's characteristic is its singularity (in terms of the time slot), but we say that an event is repeated, referring to its repeatability over the same *region*.

### 6.1 Event Movement

A TM event is intrinsically tied to the duration of time. An event may refer to a series of subevents. Dretske [22] observed that an event can move. However, it may be said that an event has moved in the sense that its TM regions have changed. Taking an example from Dretske [22], Fig. 17 shows a picnic that has moved from the building to a garden: The guests of the picnic event moved from the building to the garden. Fig. 18 shows an intermediate event (fuzzy event) during which some of the guests are in the building and some are in the garden.

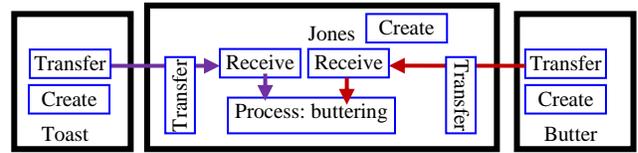

Fig. 12 The TM static representation of *Jones buttered the toast*

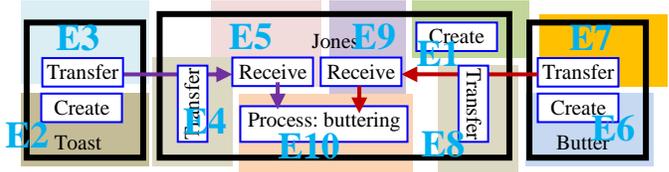

Fig. 13 The TM elementary events of *Jones buttered the toast*

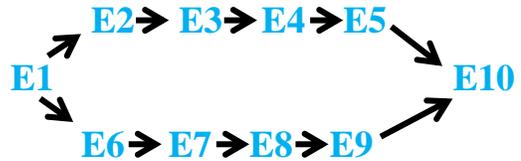

Fig. 14 The logical sequence of events of *Jones buttered the toast*

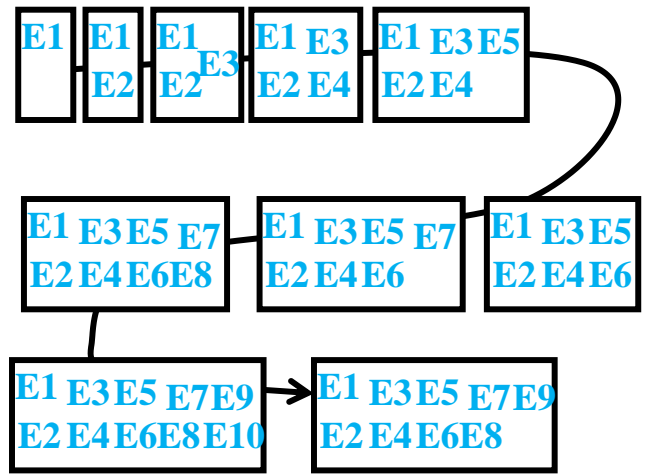

Fig. 15 Changes in the scene of *Jones buttered the toast*

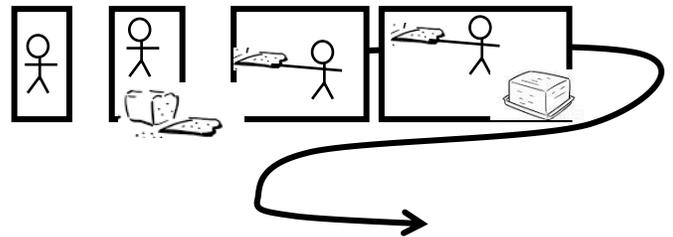

Fig. 16 A few changes in the scene of *Jones buttered the toast*

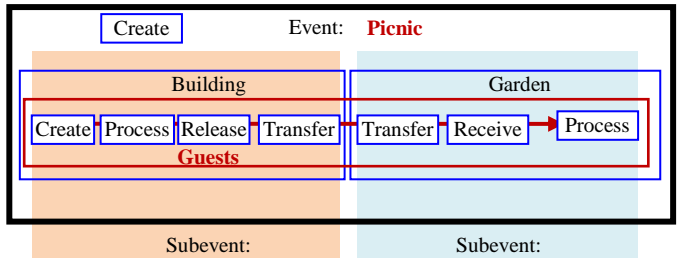

**Fig. 17 An event starts in a building and moves to the garden**

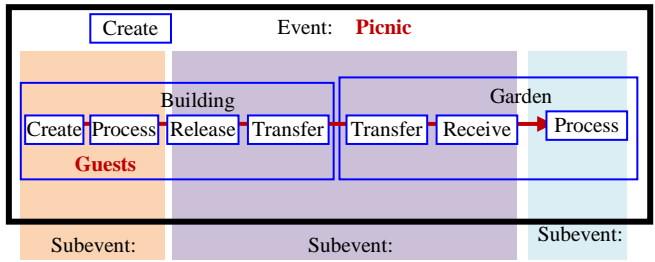

**Fig. 18 An intermediate event of a picnic that starts in a building and moves to the garden**

The classical definition is that movement (or motion) is simply a *change* in time. The physical movement is executed through infinite, continuous steps or a large number of small chemical movements (e.g., ions moving through a membrane). It is a change in a spatial position. In this context, we have to distinguish between change in a thing (TM create and TM process) and change in thimac (TM release, TH transfer, and TM receive); for example, a thing (e.g., a chameleon) may change its color when it is in the same thimac.

The nature of movements and changes in this context is worth additional study.

## 6.2 Time and Movement

Time and movement (motion) are connected to each other. Commonly, the passage of time is not, as noted, relative to the change in position. Consider Fig. 19 (left)—in a single photograph, we cannot be certain whether the dancer is moving or standing still. Observing her at different points in time, we decide that the dancer has not changed her posture in the left picture of Fig. 19.

In TM terminology, there is no change in the event region (i.e., endurance through time). According to such a view, the enduring posture is a historic route of static thimac between, say, 11:45 and 12:00, with successive time thimacs. Such a picture is similar to Whitehead's process ontology, in which objects are stable patterns of actual sequential occasions. In such an approach, change may be called a process rather than an event. According to Shipley [14], continued existence of an object is an event because it requires a reference to time. An apple falling is an event, and an apple existing in time is an event (see what happens to it after a long time of existence).

In the picture in Fig. 20 (right), there is a change (legs, hands, and head movements) to reach this second posture; hence, now, three events are illustrated in Fig. 21. The transition (the dotted V in Fig. 20) is fuzzy in the sense that it is an unstable condition, which is more than potentiality but less than actuality (event). The three events in Fig. 21 (right) have existential order (left [before] posture, [between] fuzziness, and right [after] posture).

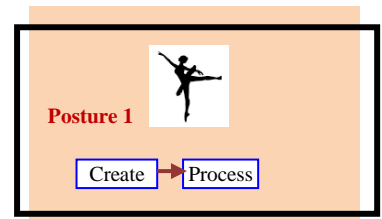

**Fig. 19 The TM model of changing posture**

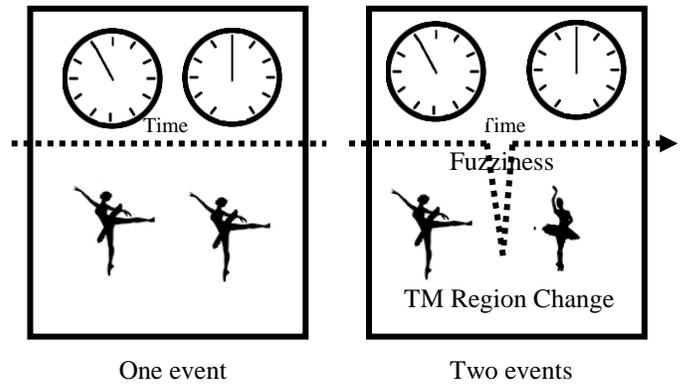

**Fig. 20 Changing posture, including the intermediate event**

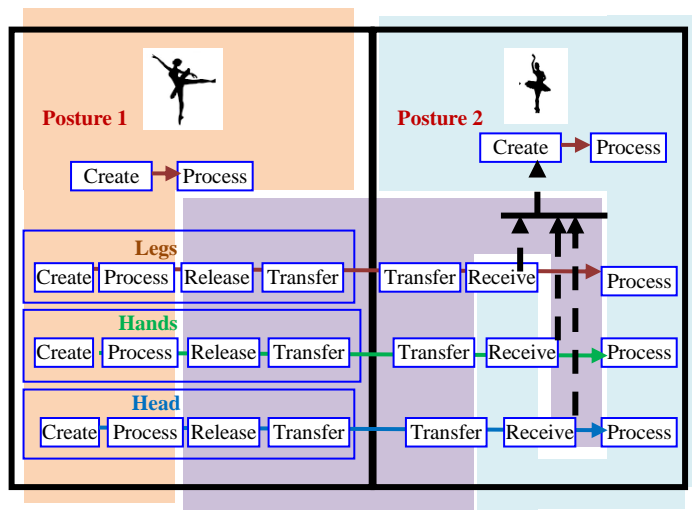

**Fig. 21 The TM model of changing posture, including the intermediate event**

## 7. Conclusion

In this paper, we further explore the TM model that concentrates on the notion of an event as a dynamic phenomenon stemming from the five generic actions in TMs. We started by giving a sample TM application for software engineering in the form of a conceptual model of a library service system. From this general applicability of the TM model, we inquired deeper into the connection between the notions of potential action in the static description and the dynamism generated by events. It is interesting that the TM model can be used in expressing a typical business process, such as a library service system, and that similarly can be utilized to model the movement of a dancer at the static and dynamic levels. This implies the model's viability as a general conceptual modeling tool.

Nevertheless, it appears that the TM modeling warrants further scrutiny and philosophical analysis.

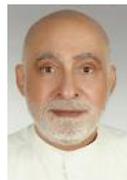

**Sabah S. Al-Fedaghi** is an associate professor in the Department of Computer Engineering at Kuwait University. He holds an MS and a PhD from the Department of Electrical Engineering and Computer Science, Northwestern University, Evanston, Illinois, and a BS from Arizona State University. He has published many journal articles and papers in conferences on software engineering, database systems, information ethics, privacy, and security. He headed the Electrical and Computer Engineering Department (1991–1994) and the Computer Engineering Department (2000–2007). He previously worked as a programmer at the Kuwait Oil Company. Dr. Al-Fedaghi has retired from the services of Kuwait University on June 2021. He is currently (Fall 2021/2022) seconded to teach in the department of computer engineering, Kuwait University.